\documentstyle[aps,preprint,epsfig]{revtex}
\begin{document}

\draft
\preprint{SNUTP 97-161}

\title{Quantum Phase Transitions and Particle-Hole Pair Transport
\\in Capacitively Coupled Josephson-Junction Chains}
\author{Mahn-Soo Choi$^1$, M.Y. Choi$^2$, and Sung-Ik Lee$^1$}
\address{$^1$ Department of Physics, Pohang University of Science and
Technology, Pohang 790-784, Korea}
\address{$^2$ Department of Physics and Center for Theoretical
Physics, Seoul National University, Seoul 151-742, Korea}
\maketitle

\thispagestyle{empty}

\begin{abstract}
We consider two chains of ultrasmall Josephson junctions, coupled
capacitively with each other, and investigate the transport of
particle-hole pairs and the quantum phase transitions at zero
temperature.  For appropriate parameter ranges, the particle-hole
pairs are found to play major roles in transport phenomena;
condensation of such pairs leads to the superconducting state,
displaying perfect drag of supercurrents along the two chains.
\end{abstract}

\pacs{PACS Numbers: 74.50.+r, 67.40.Db, 73.23.Hk}

\def\indicator{\noindent{\bf\P}\quad}
\newcommand\varD{{\cal D}}
\newcommand\varG{{\cal G}}
\newcommand\varS{{\cal S}}
\newcommand\bfq{{\bf q}}
\newcommand\bfr{{\bf r}}
\newcommand\avgl{\left\langle}
\newcommand\avgr{\right\rangle}
\newcommand\Ca{ C^{(1)} }
\newcommand\Cb{ C^{(2)} }
\newcommand\Ci{ C_I }
\newcommand\Ka{ K^{(1)} }
\newcommand\Kb{ K^{(2)} }
\newcommand\Ki{ K^{(I)} }
\newcommand\va{ v^{(1)} }
\newcommand\vb{ v^{(2)} }
\newcommand\vi{ v^{(I)} }
\newcommand\hatU{\widehat{U}}
\newcommand\hatUa{ \widehat{U}^{(1)} }
\newcommand\hatUb{ \widehat{U}^{(2)} }
\newcommand\hatUi{ \widehat{U}^{(I)} }


Both the system of small metallic tunnel junctions and that of small
Josephson junctions display the remarkable effects of Coulomb
blockade, raising the possibility of single charge (electron or Cooper
pair) tunneling~\cite{Graber92,Schonx90}.
In particular, recent theoretical predictions~\cite{Averin91a} and
experimental demonstrations~\cite{Delsin96,Matter97} with two
capacitively coupled one-dimensional (1D) arrays (i.e., chains) of
submicron metallic tunnel junctions has revealed another fascinating
role of the Coulomb interaction: The current fed through either of the
chains induces a secondary current in the other chain.
The primary and secondary currents are comparable in magnitude but
opposite in direction.
Such current drag at low bias voltage has the origin in the transport
of electron-hole pairs, which are bound by the electrostatic energy of
the coupling capacitance.
Similar current drag has also been observed in two capacitively
coupled two-dimensional (2D) electron gases~\cite{Solomo89}, although
the drag, attributed to a different mechanism~\cite{Pogreb77}, is much
smaller in strength~\cite{Nazaro97}.
On the other hand, in the Josephson-junction system, competition
between the charging energy and the Josephson coupling energy is well
known to bring quantum fluctuations, which affect in a crucial way the
phase transition of the
system~\cite{Bradle84,Faziox91,BJKim95,mschoi98a}.
Combined with such quantum fluctuation effects, the pair transport
phenomena in coupled systems, which have not been studied, are expected
to provide even richer physics.

In this Letter we consider two chains of ultrasmall Josephson
junctions, coupled capacitively with each other, and examine the
transport behaviors and quantum phase transitions at zero temperature.
For appropriate parameter ranges, the particle-hole pairs (``excitons''), 
each consisting of an excess and a deficit in Cooper
pairs across the two chains, are found to play major roles in
transport phenomena.
In particular, the condensation of such pairs leads to
the superconducting state, which is characterized by the
perfect drag of supercurrents along the two chains.

\begin{figure}
\centerline{\epsfig{file=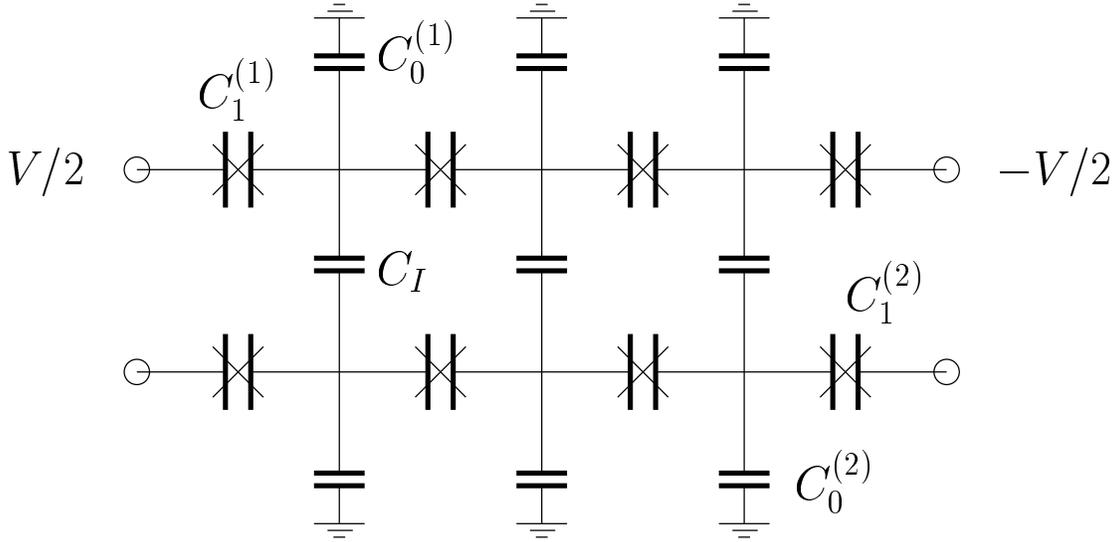,clip=,width=0.9\columnwidth}}
\caption{Schematic diagram of the system.}
\label{fig:CCJJC}
\end{figure}

We consider two chains of Josephson junctions ($a =1,2$), 
each of which is characterized
by the Josephson coupling energy $E_J$~\cite{end_note:1} and the charging
energies $E^{(a)}_0{\equiv}e^2/2C^{(a)}_0$ and
$E^{(a)}_1{\equiv}e^2/2C^{(a)}_1$, associated with the
self-capacitance $C^{(a)}_0$ and the junction capacitance $C^{(a)}_1$,
respectively (see Fig.~\ref{fig:CCJJC}).
The two chains are coupled with each other by 
capacitance $C_I$, with which the electrostatic energy
$E_I{\equiv}e^2/2\Ci$ is associated.  It is assumed that there is no
Cooper-pair tunneling between the chains~\cite{end_note:3}.  

It is convenient to write the partition function of the system in
terms of the path integral representation
\begin{equation}
Z
= \prod_{a,x,\tau}
  \sum_{n^{(a)}(x,\tau)}
  \int_0^{2\pi}\frac{d\phi^{(a)}(x,\tau)}{2\pi}
  \exp\left[ -\varS \right]
  \label{CCJJC:Z}
\end{equation}
with the Euclidean action
\begin{eqnarray}
\varS
& = & \frac{1}{2K}\sum_{a,b}\sum_{x,x',\tau}
  n^{(a)}(x,\tau)C_{ab}^{-1}(x,x')n^{(b)}(x',\tau)
  \nonumber \\
& & \mbox{}
  - K\sum_{a}\sum_{x,\tau}
    \cos\nabla_x\phi^{(a)}(x,\tau)
  \nonumber \\
& & \mbox{}
  + i\sum_{a}\sum_{x} n^{(a)}(x,\tau)
    \nabla_\tau\phi^{(a)}(x,\tau)
  \label{CCJJC:L}
  .
\end{eqnarray}
The number $n^{(a)}(x)$ of excess Cooper pairs and 
the phase $\phi^{(a)}(x)$ of
the superconducting order parameter on the grain at $x$ in
chain $a$ are quantum mechanically conjugate variables:
$[n^{(a)}(x),\phi^{(b)}(x')]{=}i\delta_{xx'}\delta_{ab}$.
All the dynamics in the system occurs over the time scale
$\omega_s^{-1}{\equiv}\hbar/\sqrt{8E_JE_s}$, where
$E_s \equiv e^2/2C_s$ with $C_s$ determined by the largest capacitances
(see below for precise determination).
Accordingly, in Eq.~(\ref{CCJJC:L}),
the coupling constant has been defined to be
$K{\equiv}\sqrt{E_J/8E_s}$, and all capacitances have been rescaled in units of
$C_s$.
Further, $\nabla_x$ and $\nabla_\tau$ denote the difference
operators with respect to the position $x$ and to the imaginary time
$\tau$, respectively, and the (imaginary) time slice $\delta\tau$
has been chosen to be unity (in units of
$\omega_s^{-1}$)~\cite{end_note:2}.  
The capacitance matrix in Eq.~(\ref{CCJJC:L}) takes the form
\begin{equation}
C
= \left( \begin{array}{cc}
    \Ca & 0 \\ 0 & \Cb
  \end{array} \right)
  + C_I\left( \begin{array}{rr}
    1 & -1 \\ -1 & 1
  \end{array} \right),
\end{equation}
where the submatrices $C^{(a)}(x,x')$ ($a{=}1,2$) are defined by the Fourier
transforms $\widetilde{C}^{(a)}(q){=}C^{(a)}_0{+}C^{(a)}_1\Delta(q)$
with $\Delta(q)\equiv2(1{-}\cos{q})$.

Following the methods in Ref.~\cite{mschoi98a}, one can transform the
system in Eqs.~(\ref{CCJJC:Z}) and (\ref{CCJJC:L}) and, apart from the
irrelevant spin-wave part, get the two coupled 2D systems of classical
vortices: 
\begin{equation}
Z_V
= \sum_{\{v\}}\exp\left[
    - 2\pi^2K\sum_{a,b} \sum_{\bfr,\bfr'}
      v^{(a)}(\bfr)U_{ab}(\bfr{-}\bfr')v^{(b)}(\bfr')
  \right]
  ,
  \label{2DVG:Z}
\end{equation}
where the 2D space-time vector notation
$\bfr \equiv (x,\tau)$ has been employed.  
The interaction $U_{ab}(\bfr)$ between
vortices can be written in the block form
\begin{equation}
U(\bfr)
= \left( \begin{array}{cc}
      U^{(1)}(\bfr) & 0 \\
      0 & U^{(2)}(\bfr)
    \end{array} \right)
  + U^{(I)}(\bfr) \left( \begin{array}{cc}
      +1 & -1 \\ -1 & +1
    \end{array} \right)
\end{equation}
where the subinteractions $U^{(a)}$ and $U^{(I)}$ are given
by the Fourier transforms
$
\widetilde{U}^{(a)}(\bfq)= 
[\widetilde{C}^{(a)}(q)\Delta(q)+D(q)\Delta(\omega)
]/T(\bfq)$ and 
$
\widetilde{U}^{(I)}(\bfq)= C_I \Delta(q)/T(\bfq)$ with
$
D(q)\equiv C_I
\widetilde{C}^{(1)}(q){+}\widetilde{C}^{(2)}(q)
] + \widetilde{C}^{(1)}(q)\widetilde{C}^{(2)}(q)$ and
$
T(\bfq) \equiv \Delta^2(q) + D(q)\Delta^2(\omega) + 
[2C_I +\widetilde{C}^{(1)}(q)+\widetilde{C}^{(2)}(q)
] \Delta(\omega)\Delta(q)$.

\begin{figure}
\centerline{\epsfig{file=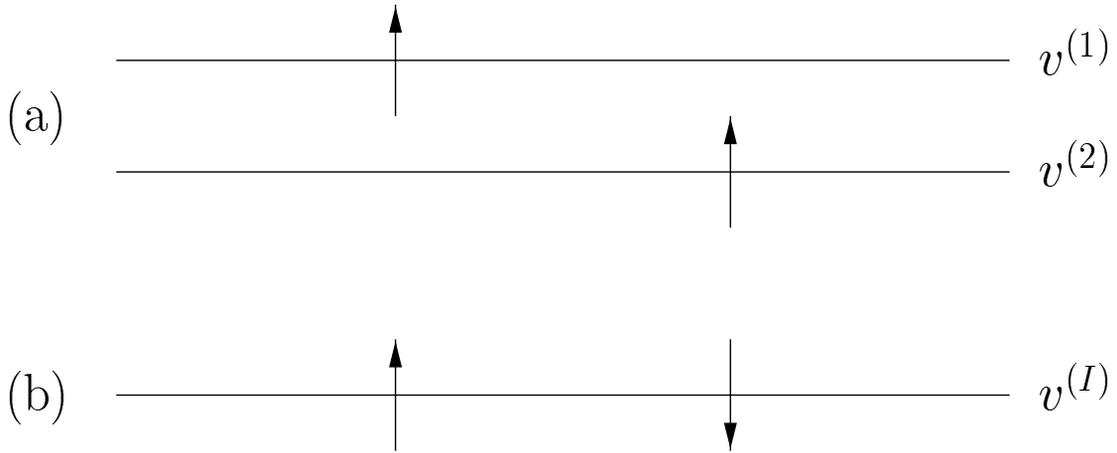,clip=,width=0.9\columnwidth}}
\caption{Displacement vortex $v^{(I)}$. (a) The configuration of one vortex
  $v^{(1)}(\bfr){=}{+}1$ on layer $1$ and another
  $v^{(2)}(\bfr'){=}{+}1$ on layer $2$ ($\bfr \neq\bfr'$) 
  corresponds to (b) a pair of displacement vortex $v^{(I)}(\bfr){=}{+}1$ 
  and antivortex $v^{(I)}(\bfr'){=}{-}1$.
  }
\label{fig:vi=v1-v2}
\end{figure}

It is of interest to express the effective model
in Eq.~(\ref{2DVG:Z}) in terms of the Hamiltonian 
\begin{eqnarray}
H_V
& = & -\pi K\sum_{a;\bfr,\bfr'}
    v^{(a)}(\bfr)\hatU^{(a)}(\bfr{-}\bfr')v^{(a)}(\bfr')
  \nonumber \\
& & -\pi K\sum_{a;\bfr,\bfr'}
    \vi(\bfr)\hatU^{(I)}(\bfr{-}\bfr')\vi(\bfr')
  \label{2DVG:H}
  ,
\end{eqnarray}
where
$
\widehat{U}^{(\alpha)}(\bfr)
\equiv 2\pi\left[U^{(\alpha)}(0)-U^{(\alpha)}(\bfr)\right]
$ ($\alpha=1,2,I$)
and 
$v^{(I)}(\bfr)\equiv \va(\bfr)-\vb(\bfr)$.
It should be noticed that for all $C_I\neq 0$,
$U^{(I)}(0)$ is always divergent and gives rise to the
{\em vortex number equality condition}
$\sum_\bfr{}v^{(1)}(\bfr) =\sum_\bfr{}v^{(2)}(\bfr)$, or equivalently,
the vorticity neutrality condition $\sum_\bfr{}v^{(I)}(\bfr) =0$ for
$v^{(I)}$.  Similar neutrality conditions $\sum_\bfr{}v^{(a)} =0$ ($a=1,2$)
should be satisfied unless $C^{(a)}_0 =0$~\cite{mschoi98a}.
In general $\widehat{U}^{(\alpha)}(\bfr)$
for $\alpha =1,2,I$ is anisotropic in the $x$ and the
$\tau$ directions, but the anisotropy is negligible at large
length scales~\cite{mschoi98a}; in the spirit of the renormalization group
analysis, $\hatU^{(\alpha)}(\bfr)$ can therefore be considered
isotropic in the space-time.
In this way, the system has reduced to three subsystems
of interacting vortices, as described by Eq.~(\ref{2DVG:H}):
$\{v^{(a)}(\bfr)\}$ on (space-time) layer $a \,(=1,2)$ and additional
vortices $\{v^{(I)}(\bfr)\}$, which
measure the relative displacements of
the vortices on the two layers. 
Figure~\ref{fig:vi=v1-v2} shows that the
configuration of the two vortices, $v^{(1)}(\bfr){=}{+}1$ on layer $1$ and
$v^{(2)}(\bfr'){=}{+}1$ on layer $2$ ($\bfr \neq\bfr'$) gives
a pair of vortex $v^{(I)}(\bfr){=}{+}1$ and antivortex
$v^{(I)}(\bfr'){=}{-}1$ for $v^{(I)}$.  For this reason, we call
$v^{(I)}$ ``displacement vortices''.
As we shall see in detail below, the displacement vortices describe 
effectively the correlations between the two chains in the system
and play a major role in phase transitions and transport
phenomena. 

The linear response $\sigma_{ab}(\omega)$ of the current
in chain $a$ to the voltage biasing chain $b$ (see
Fig.~\ref{fig:CCJJC}) can be
obtained via the analytic continuation
\begin{equation}
\sigma_{ab}(\omega)
= \frac{1}{i\omega}\lim_{q\to0}
  \widetilde\varG_{ab}(q,i\omega'\to\omega{+}i0^+),
  \label{sigma:def}
\end{equation}
where $\widetilde\varG_{ab}$ is the Fourier transform of the
imaginary-time Green's function
\begin{equation}
\varG_{ab}(x,\tau)
= \avgl T_\tau [I^{(a)}(x,\tau)I^{(b)}(0,0)] \avgr
  ,
\end{equation}
with the time-ordered product $T_\tau$
and the current operators
$I^{(a)}(x)\equiv\sin\nabla_x\phi^{(a)}(x)$.
In the same manner as that leading to Eq.~(\ref{2DVG:Z}) from
Eq.~(\ref{CCJJC:Z}), 
together with the path integral representation of the Green's
function~\cite{Bradle84}, one can obtain the expression in terms of 
vortices:
\begin{eqnarray}
&& \varG_{ab}(\bfr_1{-}\bfr_2)
= \nabla_{\tau_1}\nabla_{\tau_2}\Biggl\{
    -\frac{1}{K}U_{ab}(\bfr_1{-}\bfr_2)
    \nonumber \\
&& \mbox{}
  + 4\pi^2\sum_{a',b';\bfr_1',\bfr_2'}
    U_{aa'}(\bfr_1{-}\bfr_1')
    U_{bb'}(\bfr_2{-}\bfr_2')
    \avgl v^{(a')}(\bfr_1') v^{(b')}(\bfr_2') \avgr_V
  \Biggr\}
  ,
  \label{varG:V}
\end{eqnarray}
where $\avgl\cdots\avgr_V$ stands for the average with respect to the
effective vortex Hamiltonian in Eq.~(\ref{2DVG:H}).

Heretofore we have constructed a most general formulation for the system of
two capacitively coupled Josephson-junction chains: All the
equilibrium and transport properties of the system can be
understood through the 2D classical model in Eq.~(\ref{2DVG:Z}) or
(\ref{2DVG:H}) and the response function given by
Eqs.~(\ref{sigma:def}) and (\ref{varG:V}).
We now consider a few simple cases and investigate the interesting
phenomena associated with the particle-hole pairs.  

We first consider the case of two identical chains
($\Ca_1=\Cb_1\equiv{}C_1$) without self-capacitance
($\Ca_0=\Cb_0 =0$).  In this case, with $C_I=0$, each
chain would be always insulating regardless of the ratio of $E_1$ to
$E_J$~\cite{mschoi98a}.   
For large coupling capacitance $C_I\gg{}C_1$, the relevant capacitance
and time scales are given by $C_s=2C_I$ and
$\omega_s^{-1}=\sqrt{\hbar^2/4E_JE_I}$, respectively.
At large (space-time) length scales, the short-distance anisotropy
may be disregarded, which yields the vortex Hamiltonian 
\begin{eqnarray}
H_V
& \simeq & +\sum_a\pi K^{(a)}_{\it eff}
  \sum_{\bfr}v^{(a)}(\bfr)v^{(a)}(\bfr)
  \nonumber \\
& &
  -\pi\Ki_{\it eff}
  \sum_{\bfr\neq\bfr'}v^{(I)}(\bfr)\log|\bfr-\bfr'|v^{(I)}(\bfr')
  \label{H:I-1}
\end{eqnarray}
with the effective coupling constants
$\Ki_{\it{}eff}=\sqrt{E_J/16E_I}$ and
$\Ka_{\it{}eff}=\Kb_{\it eff}=(C_1/C_I)\Ki_{\it{}eff}$.
It should be noticed from the first term in Eq.~(\ref{H:I-1}) that
vortices on each layer does not even satisfy the vorticity neutrality
condition and would form a plasma of free vortices.
The state of the
displacement vortices $v^{(I)}$, on the other hand, depends strongly
on $\Ki_{\it{}eff}$:
For $\Ki_{\it{}eff}$ below the Berezinskii-Kosterlitz-Thouless
(BKT) transition value $K_{BKT}{\sim}2/\pi$~\cite{Berezi71}, they
can exist in free vortices; 
for $\Ki_{\it{}eff}>K_{BKT}$, only in bound dipoles.

In general the system is insulating with free vortices (induced by quantum
fluctuations)~\cite{Bradle84,mschoi98a}.
Namely, the Coulomb blockade,
disallowing single Cooper pair transport,
tends to make each separate chain insulating.
Nonetheless another transport mechanism is available 
for $\Ki_{\it{}eff}>K_{BKT}$:
To see this, we first note that the displacement vortices $\vi$ 
are nothing but topological
singularities in the 2D (space-time) configuration of the phase differences
$\phi^{(I)}\equiv\phi^{(1)}-\phi^{(2)}$.
The absence of free vortices 
for $\Ki_{\it{}eff}>K_{BKT}$ leads to the algebraic order 
of $\phi^{(I)}$ and accordingly the charge transport via
the conjugate variables $n^{(I)}\equiv (n^{(1)}{-}n^{(2)})/2$,
each representing the excess Cooper pair number difference 
across the two chains
or the number of excess-deficit pairs (i.e., particle-hole pairs).
Such transport can be confirmed by examining
the current responses given by Eq.~(\ref{sigma:def}).
Noting bound pairs of $v^{(I)}$ for $\Ki_{\it{}eff} > K_{\it{}BKT}$,
one obtains at low frequencies from Eq.~(\ref{varG:V}) 
\begin{equation}
{\rm Re\:}\sigma_{11}(\omega)
\simeq -{\rm Re\:}\sigma_{21}(\omega)
\simeq \frac{\pi}{4\Ki_{\it eff}}
  \left(2-\frac{\Ki_{\it eff}}{\Ki_R}\right)
  \delta(\omega)
  ,
  \label{tmp:a}
\end{equation}
where the renormalized coupling constant $\Ki_R$ is given
by~\cite{Josexx77}
\begin{equation}
1/\Ki_R
\equiv 1/\Ki_{\it eff}
  - \frac{\pi^2}{2}\sum_\bfr |\bfr|^2
    \avgl\vi(\bfr)\vi(0)\avgr_V
  .
  \label{Ki_R}
\end{equation}
Thus the system exhibits {\em superconductivity}
and carries currents along the two chains
{\em equally large in magnitude but opposite in direction}.
This perfect drag of supercurrents reveals that the charges 
indeed transport in the form of particle-hole pairs, which
are bound by the electrostatic energy due to $C_I$.
For $\Ki_{\it{}eff} < K_{BKT}$, on the other hand, the system
displays insulating particle-hole $IV$ characteristics,
qualitatively the same as those in Refs.~\cite{Averin91a,Delsin96,Solomo89}.
In a system with sufficiently small $C_1$ ($\Ka_{\it{}eff}\ll
\Ki_{\it{}eff}$), we therefore expect a BKT transition as $C_I$ is
increased, from the insulating ($\Ki_{\it{}eff} < K_{\it{}BKT}$) to
the superconducting state ($\Ki_{\it{}eff} > K_{\it{}BKT}$), where the
charge transport occurs in the form of particle-hole pairs.
The formation of bound dipoles of displacement vortices 
is an effective manifestation 
of the condensation of the particle-hole pairs, driving the 
superconductor-insulator transition in the system.

One also observes similar behaviors in the system of two identical chains 
without junction capacitance 
($\Ca_1 =\Cb_1 =0$ and $\Ca_0 =\Cb_0 \equiv C_0$).
$C_0$ is assumed to be small so that 
both chains, with $C_I=0$, would be in the insulating
phase~\cite{mschoi98a}.  We further assume that $C_I \gg C_0$, and
write the effective 2D vortex Hamiltonian~(\ref{2DVG:H}) in the form
\begin{eqnarray}
H_V
& \simeq & -\sum_a\pi K^{(a)}_{\it eff}
  \sum_{\bfr\neq\bfr'}v^{(a)}(\bfr)\log|\bfr-\bfr'|v^{(a)}(\bfr')
  \nonumber \\
& & \mbox{}
  - \pi\Ki_{\it eff}
  \sum_{\bfr\neq\bfr'}v^{(I)}(\bfr)\log|\bfr-\bfr'|v^{(I)}(\bfr')
  ,
  \label{tmp:e}
\end{eqnarray}
where the effective coupling constants here are given by
$\Ka_{\it{}eff}=\Kb_{\it eff}=\sqrt{E_J/8E^{(1)}_0}$ and
$\Ki_{\it{}eff}=\sqrt{E_J/16E_I}$.
Unlike the previous case, vortices $v^{(a)}$ on
each layer ($a=1,2$) as well as the displacement vortices $v^{(I)}$ should
satisfy the vorticity neutrality condition.  
Due to the condition $K^{(a)}_{\it eff} \ll K_{\it BKT}$,
vortices on each layer separately would still form a
neutral plasma of free vortices.
Increasing $C_I$, however, we expect a BKT transition 
at $\Ki_{\it{}eff} = K_{\it{}BKT}$, the superconducting
state above which is again characterized 
by the condensation of particle-hole pairs and
the current response function in Eq.~(\ref{tmp:a}). 

Finally, we consider the case $\Ca_0\ll\Ci\ll\Cb_0$
without junction capacitance ($\Ca_1 =\Cb_1 =0$),
where the relevant capacitance scale is
$C_s{=}\Cb_0$.  At large length scales, the vortex Hamiltonian is
also given by Eq.~(\ref{tmp:e}), with the effective coupling constants
$\Kb_{\it{}eff}\simeq\sqrt{E_J/8E^{(2)}_0}$,
$\Ka_{\it{}eff}\simeq\sqrt{\Ci/\Cb_0}\;\Kb_{\it{}eff}$, and
$\Ki_{\it{}eff}\simeq(\Ci/\Cb_0)\;\Kb_{\it{}eff}$.
We further assume that, with $C_I=0$, chain $1$ and chain $2$ would
be insulating and superconducting, respectively. 
This case is distinguished from the previous two in the fact that
while chain 1 is insulating [${\rm{}Re\:}\sigma_{11}(\omega)=0$],
the response of chain $2$ to the voltage biasing chain $1$
is superconducting:
\begin{equation}
{\rm Re\:}\sigma_{21}(\omega)
\simeq -\frac{\Ci}{\Cb_0}\;\frac{\pi}{\Kb_{\it eff}}
  \left(1-\frac{\Kb_{\it eff}}{\Kb_R}\right)
  \delta(\omega)
  ,
  \label{tmp:d}
\end{equation}
where $\Kb_R$ is given by Eq.~(\ref{Ki_R}) with
the superscript $I$ replaced by $2$.  This is not surprising since
the particle-hole pair, with $E_I \gg E^{(2)}_0$,
is no longer favorable.
It should also be noticed that the current response in
Eq.~(\ref{tmp:d}) is much smaller in magnitude than that to the
voltage applied across chain $2$ itself
\begin{equation}
{\rm Re\:}\sigma_{22}(\omega)
\simeq \frac{\pi}{\Kb_{\it eff}}
  \left(2-\frac{\Kb_{\it eff}}{\Kb_R}\right)
  \delta(\omega)
  .
\end{equation}
Therefore, although the particle-hole pairs still lead to 
the non-trivial current response of chain 2 to the voltage biasing chain $1$,
the quantitative effects are not so large as those in the previous cases.

In conclusion, we have investigated the properties associated with
particle-hole pairs in two capacitively coupled Josephson-junction
chains. 
In particular, the particle-hole pairs are found to play major roles
in the transport phenomena and to drive the superconductor-insulator
transition, with the superconducting state characterized by the
perfect drag of supercurrents along the two chains.
Although we have considered here the chains in the self-charging or
nearest-neighbor charging limit, the argument in Ref.~\cite{mschoi98a}
suggests that the qualitative results remain valid for realistic cases
of generic capacitances. 
Such coupled chain systems can presumably be realized in experiment by
current techniques, which have already made it possible to fabricate
submicron metallic junction arrays with large inter-array
capacitances~\cite{Delsin96,Matter97} as well as large arrays of
ultrasmall Josephson junctions~\cite{Schonx90}.
%
We also point out that quasiparticles have been safely neglected
in obtaining equilibrium properties at zero temperature.
At finite temperatures or for large voltage bias,
there may exist a significant amount of quasiparticles, which cause
dissipation in the system~\cite{Schonx90};
still in the weak-dissipation limit,
our results should not be affected qualitatively~\cite{BJKim95}.
Further, it should be kept in mind that too large voltage or current
biasing one chain can destroy the bound pairs of particles
and holes, making the supercurrent in the other chain vanish.
Finally, note the difference between the straight coupling
configuration~\cite{Matter97}, which has been studied here, and the
slanted-coupling configuration~\cite{Delsin96}: In the case of tunnel
junction arrays, the correlated sequential tunneling has been
proposed as the relevant transport mechanism in the latter, whereas in
the former the charge transport is attributed to the cotunneling
of particle-hole pairs~\cite{Averin91a}.  It would thus be of interest
to compare the two configurations in coupled Josephson-junction
systems.

We are grateful to J.~V. Jos\'e for valuable discussions and sending
us preprints.
This work was supported in part by the BSRI
Program, Ministry of Education of Korea and in part by the
Korea Science and Engineering Foundation through the SRC Program. 


\begin{thebibliography}{10}
\bibitem{end_note:1}
Identical Josephson coupling energy for both chains does not impose
any restriction here, since all the effects considered in this work
depend only on the relative strength of the Josephson coupling energy
and the charging energies.
\end{thebibliography}



\end{document}